\documentclass[12pt, fleqn]{article}

\begin{document}

\title{The meteoroid hazard for space navigation}

\author{Luigi Foschini\\
{\small Istituto FISBAT - CNR}\\
{\small Via Gobetti 101, I-40129 Bologna, Italy}\\
{\small E-mail: L.Foschini@fisbat.bo.cnr.it}}

\maketitle

\vskip 12pt
\begin{center}
Presented at the\\
\emph{Secondo Convegno Nazionale di Scienze Planetarie}\\
{\small (Second National Meeting of Planetary Sciences)}\\
Bormio (Italy), 25-31 January 1998.\\
\end{center}

\begin {abstract} 
Thanks to post-flight analyses of several artificial 
satellites carried out during last years, the meteoroids hazard for 
space navigation and in-orbit satellites permanence is now clear.  
Even if catastrophic impact is a rare event, high meteoroids fluxes 
can erode and weaken the satellite or space station main structures.  
However, the main danger seems to be the impact-generated plasma, 
which can produce electromagnetic interferences, disturbing the on-board 
electronics.
\end{abstract}

\vskip 12pt

\noindent \textbf{PACS-96.50.Kr} Meteors and meteoroids.\\
\textbf{PACS-95.40.+s} Artificial Earth satellites.

\newpage

\section{Introduction}
Till now, it was a common idea that space debris were the main threat 
for spacecrafts.  This hypotesis finds its pillar in a paper by 
Laurance and Brownlee (1986).  The work was the result of the 
\emph{Solar Max Satellite} post-flight analysis, from which was 
evident that space debris flux was several hundred times higher than 
natural meteoroids flux.

However, during last years emerge some facts that claim for a 
different reality.  During 1993 Perseids maximum, Beech and Brown 
(1993) launched an alarm, because they noted that impact probabilities 
with \emph{Hubble Space Telescope} sized objects, were low, but not 
negligible.  Really, on the night of August 12, 1993, astronauts 
onboard the \emph{Mir} Space Station reported audible meteoroids 
impacts and, then, it was verified that \emph{Mir} experienced about 
2000 hits during 24 hours and solar panels were hardly damaged (Beech 
\emph{et al.}, 1995).  In the same night, the ESA (\emph{European 
Space Agency}) lost the control of \emph{Olympus} telecommunication 
satellite.  The following investigations made clear that the failure 
was probably caused by an impact with a Perseid meteoroid 
(Caswell \emph{et al.}, 1995).

The satellites LDEF (\emph{Long Duration Exposure Facility}) and 
EURECA (\emph{EUropean REtrievable CArrier}) take the lion's share in 
these studies.  Post-flight analyses on these satellites drastically 
rescaled the theories of Laurance and Brownlee on the space debris 
danger.  McDonnell \emph{et al.} (1997b) showed that the error in the 
paper of Laurance and Brownlee was due to the use of a non correct 
formula for transformation from crater dimension to particle mass.  
Then, this scientists group elaborated a new formula, by using 
laboratory simulations (Gardner \emph{et al.}, 1997).  New fluxes 
estimates, based on data collected on EURECA and LDEF, showed that, at 
micrometer dimensions and 500 km altitude, debris population is not 
dominant as previously thought (McDonnell \emph{et al.}, 1997b).  
Really, above 30 $\mu$m ballistic limit meteoroids dominates, while, 
for thickness between 4 and 5 $\mu$m, 18\%
only of impacts are due to interplanetary matter.  Moreover, the 
debris flux does not change appreciably in the 30 $\mu$m size regime 
over the period 1980-1994, owing to atmospheric drag.

It is then very important to know the distribution and dynamics of 
interplanetary matter, in order to properly plan satellite orbits and 
space probes courses.

\section{Distribution and dynamics of interplanetary matter}
In interplanetary space, there are great matter quantities, coming 
from disruption of asteroids and comets.  The mass range is very 
wide{\footnote{Generally speaking, a `meteoroid' is intended to be a 
cometary or asteroidal body with a mass range between $10^{-9}$ and 
$10^{7}$ kg, even if there is a vivid debate around this definition.}} 
and is observable by using several techniques: visual, radar, \emph{in 
situ} and others.  Crossed references among data obtained by these 
techniques are full of difficulties, but not impossible (Ceplecha, 
1992; Foschini, 1997).

Interplanetary matter is subjected to gravitational fields of Sun and 
planets.  In a certain sense, it is possible to say that planets are 
the ``road-sweeper'' of the Solar System, because they collect all 
these particles that are in their neighbourhood for a sufficient time, 
making some gaps in meteoroids spatial distribution (\"{O}pik, 1951; 
see also the figure at p.  240 in Lindblad, 1987).  On the other hand, 
the solar radiation pressure, due to the momentum carried by solar 
photons, pushes meteoroids toward outer space.  In a heliocentric 
reference frame, let $\vec{r}$ be the radial unit vector and let 
$\vec{\theta}$ be the unit vector normal to $\vec{r}$ in the orbit 
plane; the meteoroid speed is then:

\begin{equation}
	\vec{v}= \dot{r}\vec{r}+r\dot{\theta}\vec{\theta}
	\label{}
\end{equation}

\noindent and the radiation force can be written as:

\begin{equation}
	m\dot{\vec{v}} \simeq 
	Q_{pr}(\frac{SA}{c})\left[(1-\frac{2\dot{r}}{c})\vec{r} 
	-(\frac{r\dot{\theta}}{c})\vec{\theta}\right]
	\label{}
\end{equation}

\noindent where $SA$ is the total amount of energy intercepted, per second, from 
a radiation beam of integrated flux density $S$ by a stationary, 
perfectly absorbing meteoroid of cross section $A$ and $Q_{pr}$ is the 
radiation pressure coefficient, proportional to the total momentum 
withdrawn from the beam (Burns, 1987).  Usually, the 
velocity-dependent part of (2) is called \emph{Poynting-Robertson 
effect}, while the radial term is simply called \emph{radiation 
pressure}, even if, as Burns (1987) wrote, there are also other 
accepted customs.

There is another non-gravitational force, the so-called 
\emph{Yarkovsky effect}.  This effect is connected with the name of a 
polish engineer who first described it in a pamphlet published in 
russian around 1900 (see \"{O}pik, 1951).  It is a force generated by 
asymmetries in the reradiated thermal energy from a rotating body 
exposed to Sun, because the evening hemisphere is slightly warmer than 
the morning one.  The warmer hemisphere radiated more energy, and 
hence momentum, than the other hemisphere, producing a net force, 
which depends on rotation frequency, thermal properties and dimensions 
of the body, and from Sun distance (Burns, 1987).  In a similar way, it 
is possible to speak about a ``seasonal'' effect, where the 
temperature difference between summer and winter produces a net force, 
which depends on polar axis orientation (Rubincam, 1995).  If 
collisions change the rotational state of cosmic bodies, the amplitude 
and direction of the force due to the Yarkovsky effect change 
in time in a stochastic way, producing a random walk of the major 
semiaxis of the body orbit.

The ratio among these forces can vary according to dimensions and 
masses of interplanetary bodies, distributing the matter in the Solar 
System.  Only those particles, that are in the neighbourhood of a 
planet for a little time thanks to Poynting-Robertson effect, can 
survive to the gravitational attraction (``road-sweeper'' effect).  
The Poynting-Robertson effect become fundamental for bodies in the 
range from 0.1 $\mu$m to some centimetre (Burns, 1987).  The Yarkovsky 
effect is dominant for larger bodies (0.1-100 m).  In the asteroidal 
population, this effect produces an orbital major semiaxis drift, 
driving the delivery of meteoroids toward Earth (Farinella \emph{et 
al.}, 1998).

Taking into account observations, made with several techniques, and 
the dynamics of the interplanetary matter, it is possible to 
elaborate a model, in order to know meteoroids fluxes and 
concetrations.  Now, the starting point is the model elaborated by 
Gr\"{u}n \emph{et al.} (1985) and regarding the interplanetary matter 
dynamics at 1 AU (\emph{Astronomical Unit}) from the Sun.  Throughout the 
model, the following characteristics of meteoroids population are 
considered: (1) the mean mass density is 2500 kg/m$^{3}$; (2) the 
relative speed between different meteoroids, as well as the impact 
speed on the Moon, is 20 km/s; (3) the flux onto Earth, as well as 
mutual impact flux, is isotropic.  Divine (1993) extended this model, 
taking also into account data from space probes (Pioneer 10 and 11, 
Ulysses, Galileo and Helios-1), and found five populations of 
meteoroids, each named for a distinctive characteristic of their 
orbital distributions and of their mass.  Taylor and McBride (1997) 
used data recorded by \emph{Harvard Radio Meteor Project} in order to 
extend the model by Gr\"{u}n \emph{et al.} (1985), taking into account 
the anisotropies in the meteoroids environment.  This radiant-resolved 
meteoroid model was precedeed by a reanalysis and correction of speed 
distribution obtained from radars, which can affect large meteoroids 
distribution in Divine's model (Taylor, 1995).  However, Gr\"{u}n 
\emph{et al.} (1997) recently developed a new model from the Divine's 
one, including the effect of radiation pressure on the meteoroids 
speed and considering impact directions and speeds.  The new model 
lead to four meteoroids populations on elliptical orbits and one 
moving on hyperbolic orbits.

At last, it should be pointed out that other models are now available 
(Gor'kavyi \emph{et al.}, 1997; Wasbauer \emph{et al.}, 1997).

\section{Impact probabilities}
The higher risk regions are those related to meteoroid streams, that 
when encounter the Earth give rise to meteor showers.  In these 
streams, the meteoroids number per volume unit can reach very high 
values, with geocentric speeds up to 71 km/s (\v{K}res\'{a}k, 1993; 
Jenniskens, 1995).  By the time, it is known the last meteor storm 
generated during 1966 by Leonids, when a ZHR (\emph{Zenithal Hourly 
Rate}) of 150,000 was recorded{\footnote{Jenniskens (1995) rose some 
doubts on this ZHR evaluation, that do not find a consideration in 
radar data.  According to Jenniskens, the correct value should be 
ZHR=15,000$\pm$3,000.}}.  The next perihelion passage of Leonid parent 
body, the comet P/Tempel-Tuttle, happened on February 1998.  Then a 
new meteor storm is expected.  If it will be so, there will be a 
considerable hazard for artificial satellites (Beech and Brown, 1994; 
Foschini and Cevolani, 1997).

A rough evaluation, of the impact probability of a meteoroid with an 
artificial satellite, can be made assuming an uniform and constant 
flux intercepting an area $A$ [m$^{2}$] for a time $t$ [s]:

\begin{equation}
	I=nFVAt\cdot 10^{-13}
	\label{}
\end{equation}

\noindent where $V$ [km/s] is the geocentric mean meteoroids speed and $n$ 
[km$^{-3}$] is the spatial number density of meteoroids contained, 
during normal conditions, into an equivalent 1000 km sized cube.  
During storms or outbursts, it is necessary to introduce an 
enhancement factor $F$, that for Leonids ranges from 300 to 10000 
(Beech and Brown, 1994).  In order to calculate the spatial number 
density, it is necessary to know the flux density of meteoroids $\Phi$ 
[km$^{-2}\cdot$h$^{-1}$] in a specific mass range, because $n$ 
is defined as (Koschack and Rendtel, 1990a, b):

\begin{equation}
	n=\frac{\Phi}{3600V}
	\label{}
\end{equation}

The flux evaluation is very difficult, because the Earth encounter a 
part of meteoroids only.  Moreover, it must be take into account the 
origin of meteoroids: when the parent body is a comet, a part of the 
nucleus surface only is active{\footnote{For example, when the Halley 
comet reach the Sun, about the 3\% of its surface is active.}} and, 
then, particles are expelled along several preferred directions, arranging 
themselves on slightly different orbits, even if centered along the 
orbit of the parent body.  Furthermore, considering the forces 
described in the previously section, meteoroids will move along 
particular configurations, composed by several 
filaments{\footnote{Some images of these structures can be found in: 
Beech and Brown (1994), Hughes (1995), Arter and Williams (1997).}}.  
The effect on radar and visual observations of these structures is 
that there are one or more peaks during shower activity.  Generally 
speaking, flux evaluation are made by using visual observations and 
taking into account several corrective factors, in order to offset for 
limiting elements (Koschack and Rendtel, 1990a, b; Jenniskens, 1994; 
Brown and Rendtel, 1996).  However, an important limit is due to maximum 
magnitude (+6.5), corresponding to a meteoroid of about $10^{-5}$ kg 
(Jenniskens, 1994).  But with radars, it is possible to reach 
magnitude +16, that should correspond to a mass of about $10^{-9}$ kg 
(Zhou and Kelley, 1997).  Other advantages are a wide collection area 
(for forward-scatter radars it is possible to reach some thousands of 
square kilometre) and the absence of weather, sunlight or moonlight 
limitations.  However, flux evaluations are possible imposing some 
restrictive hypoteses only: uniformity in space and time, and the mass 
index constant and equal to a mean value (Foschini, 1997).  Radar data 
allow us to make a meteoroid streams map, different from that obtained 
with visual observations.  Nevertheless, in both cases, it is possible 
to deduce that catastrophic impact is a rare event, even if when there 
is a meteor storm (Beech and Brown, 1994; Beech \emph{et al.}, 1995; 
Foschini and Cevolani, 1997).  When micrometeoroids are considered the 
impact is not so rare: for example, there is 41\% of impact
probability for a space station (1000 m$^{2}$ area, 1 hour exposure), 
with a meteoroid with mass equal or greater than $10^{-8}$ kg, if 
there will be a storm like 1966 Leonids (Foschini and Cevolani, 1997).  
Such a flux can make serious damages to mechanical structures, 
particularly solar arrays and antennas, that can not be obviously 
shielded.

\section{Impact-generated plasmas}
The \emph{Olympus} satellite experience, the post-flight analysis and 
the calculations of impact probabilities impose a revision of 
potential space dangers.  The mechanical impact do not seems to be a 
risk, as shown by several cases like the \emph{Hubble Space Telescope} 
(Herbert and McDonnell, 1997) and the \emph{Mir} space station 
(Christiansen \emph{et al.}, 1997).  However, the \emph{Olympus} 
failure is a paradigmatic example: in that case, the impact with a 
Perseid meteoroid could have generated electrical failures, leading to 
a chain reaction which culminated with an early end of the mission.  
According to Caswell \emph{et al.} (1995), a gyro motor stopped, 
probably owing to a lack of power, and the satellite lost the 
reference.  Following manoeuvres in order to acquire a new reference 
(\emph{Emergency Sun Acquisition}) failed, probably owing to a short 
circuit in a capacitor of the emergency network.  Even if there is not 
any certainty, it seems that the impact of a small meteoroid may have 
generated a plasma triggering discharge of charged surfaces, entering 
the grounded spacecraft via the umbilical.

After the \emph{Olympus} end-of-life anomaly, other authors looked to 
impact-produced plasma, rather than the impact itself (Brown \emph{et 
al.}, 1996; McDonnell \emph{et al.}, 1997a).  It is well known that, 
during an hypervelocity impact, a fraction of the projectile and 
target materials is evaporated and even ionized (Fechtig \emph{et 
al.}, 1978).  A plasma cloud is then created almost instantaneously 
after the impact and expands into the surrounding vacuum.  McDonnell 
\emph{et al.} (1997a) find an empirical formula for evaluation of 
charge production during an hypervelocity impact.  This equation, 
rearranged in order to emphasize projectiles dimensions and densities, 
can be written as:

\begin{equation}
	Q\simeq 3.04 \delta^{1.02}r^{3.06}V^{3.48} \ [\mathrm{C}]
	\label{}
\end{equation}

\noindent where $\delta$ is the meteoroid density [kg/m$^{3}$], $r$ is the 
meteoroid radius [m] and $V$ its speed [km/s].

Because of the energy range, the plasma production is related to 
chemical composition of meteoroid.  Cometary streams, richer of low 
ionization potential elements, will be more dangerous than other.  The 
Leonid meteoroid stream results to be the most dangerous stream, even 
during normal condition (McDonnell \emph{et al.}, 1997a).

The impact-produced plasma can disturb the satellite in several ways: 
if directly injected into circuits can destroy part of the 
onboard electronics (McDonnell \emph{et al.}, 1997a); thermal forces 
can magnetize the neighbourhood of craters (Cerroni and Martelli, 
1982); electromagnetic radiations emitted from the plasma can disturb 
several resources or scientific experiments on the satellite 
(Foschini, 1998).  Even if satellites are actually submitted to 
several procedures for electromagnetic compatibility (EMC), meteoroid 
impacts call for new studies on these arguments.  For example, the 
plasma-generated charge can deposit on near surfaces and, subsequently 
discharges to mass.  The pulse shape will depend on electric 
characteristics (resistance, inductivity, capacity) of employed 
materials.  Moreover, the pulse can disturb the onboard electronics, 
mainly in four ways (Audone, 1993; Foschini and Gallerani, 1993):

\begin{enumerate}
\item the discharge can be directly injected into a circuit; 

\item the discharge can hit a nearly surface and disturb a circuit by a 
secondary discharge; 

\item capacitive coupling between the discharge electric field and the circuit; 

\item inductive coupling between the discharge magnetic field and the circuit.
\end{enumerate}

If the first two modes are localized, and then depend on the impact 
place, the third and fourth modes can disturb distant components.  
However, these coupling effects are strongly non-linear and depending 
on circuit layout.  Thus, more detailed studies can be made on specific 
satellite only.

\section{Conclusions}
The threat from meteoroids must be revised, taking into account 
experiences such as \emph{Olympus} and experimental studies (McDonnell 
\emph{et al.}, 1997a).  Mechanical damages are localized, sporadically 
hit important parts and the catastrophic impact is an event still 
rare.  On the other hand, if the plasma charge and current production 
are considered, then the risk increase and meteoroid streams, 
particularly those composed with low ionization potential elements, 
can be dangerous even during normal conditions.  More studies about 
electromagnetic interferences from impact-produced plasmas are 
required.

\section{Acknowledgements}
Author wishes to thanks Paolo Farinella, of the Department of 
Mathematics of the University of Pisa, for constructive review.

\section{References}
Arter T.R. and Williams I.P.: 1997, ``Periodic behaviour of the April 
Lyrids.''  \emph{Mon.  Not.  R. Astron.  Soc.}, \textbf{286}, 163-172.

Audone B.: 1993, \emph{Compatibilit\`{a} elettromagnetica - 
Interferenza e immunit\`{a} di apparati e sistemi}, McGraw-Hill, 
Milano.

Beech M. and Brown P.: 1993, ``Impact probabilities on artificial 
satellites for the 1993 Perseid meteoroid stream.''  \emph{Mon.  
Not.  R. Astron.  Soc.}, \textbf{262}, L35-L36.

Beech M. and Brown P.: 1994, ``Space-Platform impact probabilities - 
The threat of the Leonids.''  \emph{ESA J.}, \textbf{18}, 63-72.

Beech M., Brown P. and Jones J.: 1995, ``The potential danger to space 
platforms from meteor storm activity.''  \emph{Quart.  J. R. 
Astron.  Soc.}, \textbf{36}, 127-152.

Brown P. and Rendtel J.: 1996, ``The Perseid meteoroid stream: 
characterization of recent activity from visual observations.''  
\emph{Icarus}, \textbf{124}, 414-428.

Brown P., Jones J. and Beech M.: 1996, ``The danger to satellites from 
meteor storms - A case study of the Leonids.''  \emph{Proc.  
5$^{th}$ Int.  Conf.  on Space}, 13-19, American Society of Civil 
Engineers.

Burns J.A.: 1987, ``The motion of interplanetary dust.''  
\emph{Proc.  Int.  School Phys.  ``Enrico Fermi'' - Course XCVIII - 
The evolution of the small bodies of the Solar System} (M. Fulchignoni 
e \v{L}.  Kres\'{a}k Eds.), 252-275, North-Holland.

Caswell R.D., McBride N. and Taylor A.: 1995, ``Olympus end of life 
anomaly - A Perseid meteoroid impact event?.''  \emph{Int.  J. 
Impact Eng.}, \textbf{17}, 139-150.

Ceplecha Z.: 1992, ``Influx of interplanetary bodies onto Earth.''  
\emph{Astron.  Astrophys.}, \textbf{263}, 361-366.

Cerroni P. and Martelli G.: 1982, ``Magnification of pre-existing 
magnetic fields in impact-produced plasmas, with reference to impact 
craters.''  \emph{Planet.  Space Sci.}, \textbf{30}, 395-398.

Christiansen E.J., Hyde J.L. and Lear D.: 1997, ``Meteoroid/orbital 
debris impact damage predictions for the Russian space station Mir.''  
\emph{Proc.  2$^{nd}$ Eur.  Conf.  Space Debris}, 503-508, 
ESA-ESOC.

Divine N.: 1993, ``Five populations of interplanetary meteoroids.''  
\emph{J. Geophys.  Res.  - Planets}, \textbf{98}, 17029-17048.

Farinella P., Vokrouhlick\'{y} D. and Hartmann W.K.: 1998, ``Meteorite 
delivery via Yarkovsky orbital drift.''  \emph{Icarus}, \textbf{132}, 378-387.

Fechtig H., Gr\"{u}n E. and Kissel J.: 1978, ``Laboratory 
simulation.''  \emph{Cosmic dust} (J.A.M. McDonnell Ed.), 607-669, 
J. Wiley \& Sons.

Foschini L.: 1997, ``On radar measurements of the terrestrial mass 
accretion rate.''  \emph{Nuovo Cimento C}, \textbf{20}, 127-130.

Foschini L.: 1998, ``Electromagnetic interferences from plasmas 
generated in meteoroids impacts.'' \emph{Europhys. Lett.}, \textbf{43}, 226-229.

Foschini L. and Cevolani G.: 1997, ``Impact probabilities of meteoroid 
streams with artificial satellites: an assessment.''  \emph{Nuovo 
Cimento C}, \textbf{20}, 211-215.

Foschini L. and Gallerani A.: 1993, \emph{Introduzione alla 
compatibilit\`{a} elettromagnetica}, Istituto di Elettrotecnica, 
Universit\`{a} di Bologna, Internal Report 2/1993, Bologna.

Gardner D.J., Shrine N.R.G. and McDonnell J.A.M.: 1997, 
``Determination of hypervelocity impactor size from thin target 
spacecraft prenetrations.''  \emph{Proc.  2$^{nd}$ Eur.  Conf.  
Space Debris}, 493-496, ESA-ESOC.

Gor'kavyi N.N., Ozernoy L.M. and Mather J.C.: 1997, ``A new approach 
to dynamical evolution of interplanetary dust.''  \emph{Astrophys.  
J.}, \textbf{474}, 496-502.

Gr\"{u}n E., Zook H.A., Fechtig H. and Giese R.H.: 1985, ``Collisional 
balance of the meteoritic complex.''  \emph{Icarus}, \textbf{62}, 
244-272.

Gr\"{u}n E., Staubach P., Baguhl M., Hamilton D.P., Zook H.A., Dermott 
S., Gustafson B.A., Fechtig H., Kissel J., Linkert D., Linkert G., 
Srama R., Hanner M.S., Polanskey C., Horanyi M., Lindblad B.A., Mann 
I., McDonnell J.A.M., Morfill G.E. and Schwehm G.: 1997, ``South-North 
and radial traverses through the interplanetary dust cloud.''  
\emph{Icarus}, \textbf{129}, 270-288.

Herbert M.K. and McDonnell J.A.M.: 1997, ``Morphological 
classification of impacts on the EURECA \& Hubble Space Telescope 
solar arrays.''  \emph{Proc.  2$^{nd}$ Eur.  Conf.  Space Debris}, 
169-175, ESA-ESOC.

Hughes D.W.: 1995, ``The Perseid meteor shower.''  \emph{Earth, 
Moon, Planets}, \textbf{68}, 31-70.

Jenniskens P.: 1994, ``Meteor stream activity I - The annual 
streams.''  \emph{Astron.  Astrophys.}, \textbf{287}, 990-1013.

Jenniskens P.: 1995, ``Meteor stream activity II - Meteor outbursts.''  
\emph{Astron.  Astrophys.}, \textbf{295}, 206-235.

Koschack R. and Rendtel J.: 1990a, ``Determination of spatial number 
density and mass index from visual meteor observations (I).''  
\emph{WGN J. Int.  Meteor Org.}, \textbf{18}, 44-58.

Koschack R. and Rendtel J.: 1990b, ``Determination of spatial number 
density and mass index from visual meteor observations (II).''  
\emph{WGN J. Int.  Meteor Org.}, \textbf{18}, 119-140.

\v{K}res\'{a}k L.: 1993, ``Cometary dust trails and meteor storms.''  
\emph{Astron.  Astrophys.}, \textbf{279}, 646-660.

Laurance M.R. and Brownlee D.E.: 1986, ``The flux of meteoroids and 
orbital space debris striking satellites in low earth orbit.''  
\emph{Nature}, \textbf{323},136-138.

Lindblad B.A.: 1987, ``Physics and orbits of meteoroids.''  
\emph{Proc.  Int.  School Phys.  ``Enrico Fermi'' - Course XCVIII - 
The evolution of the small bodies of the Solar System} (M. Fulchignoni 
e \v{L}.  Kres\'{a}k Eds.), 229-251, North-Holland.

McDonnell J.A.M., McBride N. and Gardner D.J.: 1997a, ``The Leonid 
meteoroid stream: spacecraft interactions and effects.''  
\emph{Proc.  2$^{nd}$ Eur.  Conf.  Space Debris}, 391-396, 
ESA-ESOC.

McDonnell J.A.M., Ratcliffe P.R., Green S.F., McBride N. and Collier 
I.: 1997b, ``Microparticle populations at LEO altitudes: recent 
spacecraft measurements.''  \emph{Icarus}, \textbf{127}, 55-64.

\"{O}pik E.J.: 1951, ``Collision probabilities with the planets and 
the distribution of interplanetary matter.''  \emph{Proc.  R. Irish 
Acad.  Sect.  A}, \textbf{54}, 165-199.

Rubincam D.P.: 1995, ``Asteroid orbit evolution due to thermal drag.''  
\emph{J. Geoph.  Res.  - Planets}, \textbf{100}, 1585-1594.

Taylor A.D.: 1995, ``The Harvard Radio Meteor Project meteor velocity 
distribution reappraised.''  \emph{Icarus}, \textbf{116}, 154-158.

Taylor A.D. and McBride N.: 1997, ``A radiant-resolved meteoroid 
model.''  \emph{Proc.  2$^{nd}$ Eur.  Conf.  Space Debris}, 
375-380, ESA-ESOC.

Wasbauer J.J., Blanc M., Alby F. and Ch\`{e}oux-Damas P.: 1997, 
``Modeling interplanetary dust distribution.''  \emph{Proc.  
2$^{nd}$ Eur.  Conf.  Space Debris}, 381-386, ESA-ESOC.

Zhou Q.H. and Kelley M.C.: 1997, ``Meteor observations by the 
Arecibo 430 MHz incoherent scatter radar.  II. Results from 
time-resolved observations.''  \emph{J. Atmos.  Terr.  Phys.}, 
\textbf{59}, 739-752.

\end{document}